\documentstyle{elsart}
\begin{document}
\phantom{..} \hspace{9.2 cm}FZJ--IKP(TH)--1998--31 \\
\phantom{.} \hspace{9 cm}NT@UW--99--40 \\
\phantom{.} \hspace{9 cm}DOE/ER/40561--63--INT99
\begin{frontmatter}

\title{On momentum dependence of the reaction  $\pi^- p \to \omega n$ near
threshold}

\author{C.Hanhart$^{a,b}$, A.Kudryavtsev$^{b,c}$}

\small{$^a$Department of Physics and INT, University of Washington, \\
Seattle, WA 98195, USA} \\
\small{$^b$Institut f\"{u}r Kernphysik, Forschungszentrum J\"{u}lich
GmbH,}\\
\small{ D--52425 J\"{u}lich, Germany} \\
\small{$^c$Institute of Theoretical and Experimental Physics,} \\
\small{117258, B.Cheremushkinskaya 25, Moscow, Russia}

\begin{abstract}
We discuss the near--threshold behavior
of the $\omega$ production amplitude in the reaction
$\pi^-p \to \omega n$.
In contrast to the results of earlier analyses
we find that the averaged
 squared matrix element
of the production amplitude must be a decreasing function of energy
in order to describe
the existing experimental data.
\end{abstract}
\end{frontmatter}

\section{Introduction}
\label{sec:intro}
The reaction $\pi^-p \to \omega n$  near threshold was studied
relatively long ago \cite{Bin,Key,Kar}.
The authors of those papers claimed to have found an abnormal behavior of the
production amplitude for this reaction near threshold
that is not yet understood theoretically \cite{wil}.
This conclusion was based on a comparison of the measured
cross section with that for the production of a stable particle.
Specifically, they found that the production cross
section is proportional to $P^* \, ^2$ instead of
$P^*$, as expected.
($P^*$ denotes the momentum of the outgoing neutron in the
center of momentum system.)
This behavior was interpreted as possible evidence
for a resonance in the $\omega N$ system not far above
threshold \cite{Key}.  At the same time there are no
direct indications of the existence of such a resonance
in the $\pi^-p$--channel.  Recently a behavior similar to
that of the cross section under discussion was also found
in the reaction $pd \to \omega \, ^3He$
\cite{Wurz}.

The transition amplitude $\pi N \to \omega N$, while interesting in its
own right, is also of great importance in other reactions. Theoretical
analyses of the reactions $pp \to pp\omega$ \cite{wil},
$pn \to d\omega$ \cite{kon} and $dp \to {}^3He\omega$ \cite{tobechecked},
as well as $\omega$ production in proton--nucleus collisions \cite{sib}
all rely on the $\pi N \to \omega N$ transition amplitude, in which the
pion enters as an exchanged particle, as the basic mechanism for the reactions
studied.  It is thus of great theoretical interest to obtain direct, reliable
experimental information on this reaction.

\cite{tobechecked}
information

The experiments cited above were all performed in an unusual kinematical
situation: instead of measuring the momentum distribution of the final
state for a fixed beam energy, the excitation function for a fixed neutron
momentum versus initial
energy was measured.

In this work we analyze the general expression for the production
cross section of unstable
particles in near--threshold binary reactions.
Both situations -- the standard one, wherein the energy is fixed and the final
state
momentum is varied -- and  that of the above experiments, wherein one of the
final momenta is fixed while  varying the energy, are compared.
We shall demonstrate that the dependence of the count rates on the
outgoing center of mass momentum depends on how the analysis is done.
We conclude  that the behavior of the $\omega n$ amplitude
is quite  normal. That means  that the earlier interpretation of the
experimental
data \cite{Bin,Key,Kar,Wurz} is incorrect.
To explain the results of the cited papers, we need a smooth behavior of the
averaged matrix element that must be a decreasing function of energy
in the near--threshold region.

We begin by discussing the production of a resonance
using a  monochromatic beam. We then consider the integration of the
so-obtained cross section over the beam energy, which is the procedure that was
carried out experimentally in refs.  \cite{Bin,Key,Kar}. We close with a
discussion of the formulae used in the cited papers to analyze the
experimental data.

\section{ Cross section for the production of an unstable particle}
\label{sec:stab}

Let us consider the case of a monochromatic beam of energy $E$\footnote{Note,
that all kinematical quantities are given in the center of mass.}. In this case
 the
differential cross section $d\sigma/d\Omega$ for production of a stable
particle
is
\begin{eqnarray}
\frac{d\sigma}{d\Omega}\left| \frac{}{}_{stable}\right. =\frac{\mu_i(2\pi)^4}
{p_i}\int \mid T(E,\vec
k)\mid^2
\delta(E-M-m-k^2/2\mu )k^2 dk,
\label{stabl}
\end{eqnarray}
where $\mu_i$ ($\mu$) and $p_i$ ($k$) denote the reduced mass and the
 momentum of the initial state (final state)
respectively.
This integral is proportional to $k(E,m)=\sqrt{2\mu(E-M-m)}$ after the
integration is
performed. If we consider the cross section for the  production of an omega
meson
 or any other
resonance with finite width $\Gamma$,  expression
(\ref{stabl}) must be convoluted
with the spectral density $\rho(m,\Gamma)$.
For simplicity we use a Breit-Wigner form for the spectral density,
namely
\begin{eqnarray}
\rho(m,\Gamma )=\frac{\Gamma/2\pi}{(m-\bar m)^2+\Gamma^2/4} \ .
\label{bw}
\end{eqnarray}
Here  $\bar m$ is the average mass of the unstable particle.
In this case the resonance production cross section is given by the expression:
\begin{eqnarray}
\nonumber
\frac{d\sigma}{d\Omega}\left| \frac{}{}_{unstable}\right.
&=&\frac{\mu_i(2\pi)^4}{p_i} \\
&\times& \int_{0}^{K_{max}}\frac
{ \Gamma/2\pi}{(E_{kin}-k^2/2\mu)^2+\Gamma^2/4}\mid T(E,\vec k)\mid^2 k^2dk \ ,
\label{invo}
\end{eqnarray}
where $E_{kin}=E-M-\bar m$ and $K_{max}$ is the maximum momentum of the
outgoing
neutron for the
 reaction $\pi^-p \to \omega n$. $K_{max}$ is determined by the masses
of the lightest decay products of the unstable particle.

Note that, because of the experimental setup, the authors of the papers [1--3]
have  not measured the total
 differential cross section for the  production of an unstable particle as
given by eq. (3), but only a fraction of
it, as the momentum of the outgoing neutron was constrained to lie
in a small band around a given $P^*$. In
other words, they have measured the following part of differential cross
section:
\begin{eqnarray}
\nonumber
\frac{d\sigma}{d\Omega}
\left|\frac{}{}_{\Delta P}\right.
&=&\frac{\mu_i(2\pi)^4}{p_i} \\
&\times& \int_{P^*-\Delta P/2}^{P^*+\Delta P/2}\frac
{ \Gamma/2\pi}{(E_{kin}-k^2/2\mu)^2+\Gamma^2/4}\mid T(E,\vec k)\mid^2 k^2dk \ ,
\label{invo1}
\end{eqnarray}
Let us estimate the remaining integral
for the case when the scattering amplitude $\mid T(E,\vec k)\mid$
is approximately constant in the interval $\Delta P$, as
one would expect close to the production threshold.
In this case we get
\begin{eqnarray}
\frac {d\sigma}{d\Omega}\left| \frac{}{}_{\Delta P}\right. \propto
\mid T(E,P^*)\mid^2 I(E_{kin}),
\end{eqnarray}
where
\begin{eqnarray}
I(E_{kin})=\frac{\mu}{\pi}\sqrt{\mu\Gamma}\int_{a_-}^{a_+} \frac{ \sqrt y
dy}{(y-y_0)^2+1}
\label{integr}
\end{eqnarray}
and $y_0=2E_{kin}/\Gamma$.
Here the limits are
$a_\pm=\frac{(P^*\pm \Delta P/2)^2}{\mu\Gamma}$\ .
As will become clear below, the behavior of the integral depends on the
parameter
\begin{eqnarray}
\chi (P^*) := a_+-a_- \equiv \frac{2P^*\Delta P}{\mu\Gamma} \ .
\label{chidef}
\end{eqnarray}

Let us consider the case of small $P^*$ and  $\Delta P$ such that the condition
\begin{eqnarray}
I(E_{kin})\approx
\chi(P^*) \ll 1 \
\label{chi}
\end{eqnarray}
is satisfied. In this limit the denominator
under the integral is practically constant, so that we have
\begin{eqnarray}
I(E_{kin})\approx
\frac{\mu}{\pi}\sqrt{\mu\Gamma}\frac{1}{(y^*-y_0)^2+1}\int_{a_-}^{a_+}\sqrt
ydy=
\frac{2}{\pi\Gamma}\frac{P^{*2}\Delta P+\frac{1}{12}(\Delta P)^3}{(y^*-y_0)^2+1}\,,
\label{wide}
\end{eqnarray}
where $y^*=P^*{}^{2}/\mu\Gamma$.
The dependence of this integral on energy looks like a BW-resonance with
strength
proportional to $P^{*2}$. This was the dependence found in ref. \cite{Bin}.

\begin{figure}[h]
\vspace{12cm}
\includegraphics{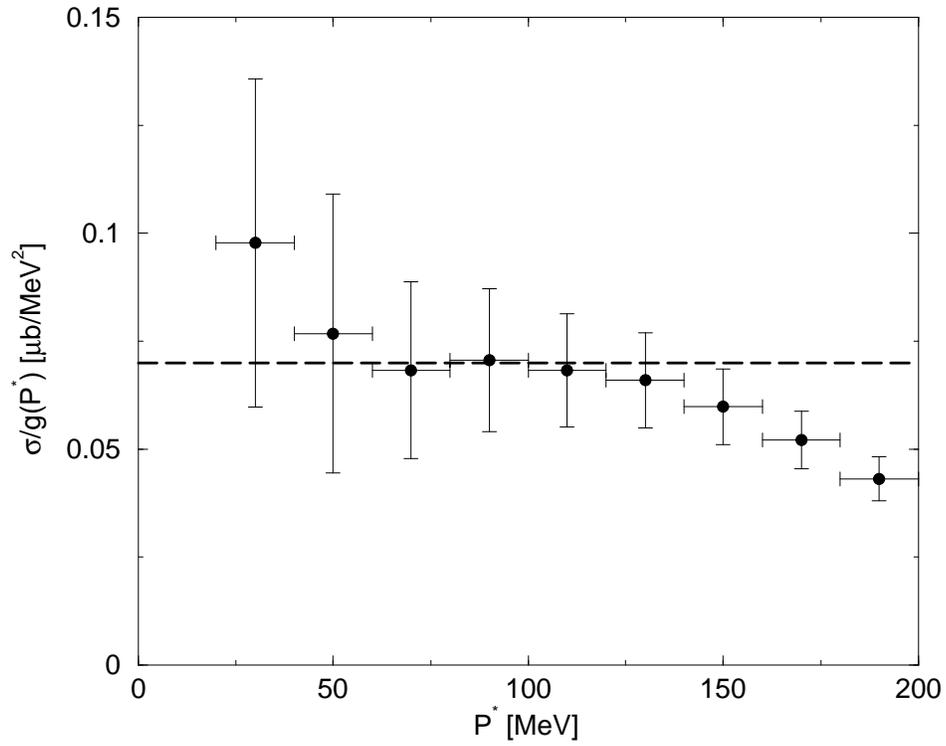}
\caption{\it{The cross section data normalized to
$g(P^*) = P^* {}^2 + \frac{1}{12}(\Delta P)^2$. The curve is introduced to guide
the eye. The data are from refs. \protect{\cite{Key,Kar}},
where the errors where modified according to the uncertainty in $P^*$.}}
\label{fig}\end{figure}

\label{sec:integr}
In the experiments \cite{Bin,Key,Kar} an additional integration over the
beam energy (still keeping $P^*$ fixed) was performed in order to remove
the width-dependence
from eq. (\ref{wide}).  Indeed, since the spectral density is normalized,
integrating
over the beam energy gives
\begin{eqnarray}
\int dE_{kin} I(E_{kin}) \simeq P^* {}^2 \Delta P + \frac{1}{12}(\Delta P)^3 \ .
\label{quadrat}
\end{eqnarray}
The above derivation shows specifically that
\begin {eqnarray}
\mid T(E,P^*)\mid^2 \propto \frac{1}{g(P^*)}
\frac{d\sigma}{d\Omega}\left|
\frac{}{}_{\Delta P}\right. \ ,
\label{tqadrat}
\end{eqnarray}
where $g(P^*) = P^* {}^2 + \frac{1}{12}(\Delta P)^2$.
The right hand side of this equation is displayed in figure \ref{fig}, based
on the data of ref. \cite{Kar} for the points with $P^* \ge 50 MeV/c$.
As for the point at
  $30 MeV/c $, we used the data from ref. \cite {Key}, averaged over an
interval of
 $\Delta P=20 MeV/c$.
Note that the errors of $(d\sigma / P^*{}^2)$ displayed in the plot contain
 the uncertainty in $P^*$ as well.
 Figure 1 gives evidence for a matrix
 element  $\mid T(E,P^*)\mid $ that is practically constant,
 at least for the points $P^* \le  110 MeV/c $.
The data above  $110 MeV/c$ give evidence for
a matrix element that decreases smoothly with $P^*$, as would
be expected in the usual effective range approximation.
We conclude therefore that the existing experimental data
\cite{Bin,Key,Kar} for the reaction $\pi^-p\to \omega  n$ give no indication
of a growth  of the matrix element for increasing $P^*$ in a wide interval
of momenta $P^*$
above threshold.
Note that the over--all dependence of the matrix element on $P^*$ is
contrary to the conclusions of refs. [1-3].

\label{sec:chi}

We now investigate the  second limiting case,

\begin{eqnarray}
\chi(P^*)\gg1.
\label{more}
\end{eqnarray}
To estimate the  integral  (\ref{integr}) in this situation
we must further distinguish separately two possibilities:

i) The
energy
parameter $y_0$ is
within the limits $a_+$ and $a_-$ of the
integral (\ref{integr}). In this case
\begin {eqnarray}
 I(E_{kin})\approx\sqrt{2\mu E_{kin}}.
\label{pstar}
\end{eqnarray}

ii) The energy parameter $y_0$ is not within the interval $[a_-,a_+]$.
The integral $I(E_{kin})$ is then strongly suppressed.

In short, if condition (\ref {more}) is satisfied we get the usual energy
behavior for
the differential cross section, namely a linear $P^*$--dependence.

The condition
$$ \chi(P^*) \approx 1$$
determines the critical value of $P^*$.  Thus, by measuring the count rates
 versus
$P^*$
one may observe a transition from a $P^{*2}$ behavior of the cross section at
low $P^*$
to
a linear dependence at high $P^*$, even for a constant matrix element.
 In the case of the omega this takes place
at  $P^*_{cr}=\mu\Gamma/2\Delta P \approx 90 MeV/c$, if
$\Delta P \approx 20$ MeV, as specified in ref. \cite{Key}.

In ref. \cite{Bin} the production of $\eta$ and $\eta\prime$
 was studied as well. The authors
report that here a behavior very different from what they found for
$\omega$ production.
Using the above discussion one can now easily understand this: for both mesons
condition
(\ref {more}) was satisfied, since $P^*_{cr}=0.01$ MeV for the $\eta$ and
$P^*_{cr}=2.4$ MeV
for the $\eta\prime$.

\label{sec:incor}

To complete our criticism of the analyses of refs. \cite{Bin,Key,Kar}, we
compare our formulae to those given therein. Let us start by briefly repeating
the
arguments for a linear dependence of the $\omega$ production cross section on
$P^*$
 given in \cite{Bin}.
Instead of eq. (1) for the production of a stable particle, ref. \cite{Bin}
starts directly from
the expression for the double--differential cross section,
\begin{equation}
\frac {d^2\sigma}{dmd\Omega} \propto \rho (m,\Gamma) P^* \ .
\label{pru}
\end{equation}
 In order to get the total production rates  for producing
 final particles with a given  $P^*$,  expression (\ref{pru}) was
integrated over
the initial
 energy under the constraint $P^*=const$. By employing energy conservation,
i.e. using the condition
\begin{equation}
 dm = dE \ \  \mbox{for} \ \  P^* \
\mbox{fixed}  \ ,
\label{encon}
\end{equation}
(\ref{pru}) can be formally integrated.
Since the spectral density is normalized,
this integration yields
\begin{equation}
\frac{d\sigma (P^*)}{d\Omega} \propto P^* \ .
\label{binniecs}
\end{equation}
This procedure to obtain  eq. (\ref{binniecs}) from eq. (\ref{pru}) looks
formally correct,
but it is not. The reason for this is that in order to derive eq. (\ref{pru})
the
energy conserving $\delta$--function in eq. (1) was  evaluated. Therefore $P^*$
in
eq.  (\ref{pru}) implicitly depends on $E$ and $m$ and thus must be
treated as  a dependent variable
in any argument based on eq. (\ref{pru}).  Therefore the use of the relation
(\ref{encon}) in this context is simply incorrect. Instead, the
condition of allowing $P^*$ to vary only in a small interval translates into a
condition
on the ranges of integration of $m$, given a fixed energy $E$, namely
\begin{equation}
\frac {d\sigma}{d\Omega} \propto \int_{m_0-\chi \Gamma  / 2}^{m_0+\chi \Gamma
 / 2} \rho (m,\Gamma) P^* \ ,
\label{invo2}
\end{equation}
with $m_0 = E-M-P^*{}^2/2\mu$. This formula actually agrees with our eq. (5) if
we
rewrite it
in terms of an integration over $dm$.
Therefore we conclude
that in the theoretical analysis of
refs. \cite{Bin,Key,Kar} the limits of integration were not properly treated,
thereby leading to
an inappropriate conclusion for the momentum dependence of the cross section.

 It is this point that was overlooked
in the earlier works. If we impose the limit $\Delta P \to 0$ on
eq. (\ref{invo2}) and integrate over the energy we again find
$$
\frac{d\sigma}{d\Omega} \propto P^* {}^2 \Delta P  \ .
$$

\begin{figure}[h]
\vspace{5cm}
\includegraphics{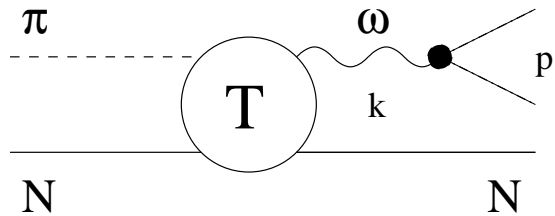}
\caption{\it{Illustration of the reaction $\pi N \to \omega N$ as
a three particle reaction taking into account the decay of the $\omega$.}}
\label{diagram}
\end{figure}

To clarify the situation we would like to add that the final result agrees
with what one expects when taking the decay of the unstable particle into
account explicitly.
Let us, for simplicity, assume a two particle decay\footnote{What follows
is exact under the assumption that the unstable particle decays
into this channel only. However, the generalization is straightforward
and only complicates the argument.},  as  illustrated
in figure \ref{diagram}. In this case the phase
space is the 3 body phase space and we get
\begin{equation}
d\sigma \propto d^3k  d\Omega_p \mu_d |\vec p|
 \mid T(E,\vec k) D_\omega (P^2)
W(m,p)\mid ^2 \ ,
\label{3part}
\end{equation}
where $D_\omega$ denotes the dressed $\omega$ propagator, $W$ is the
decay amplitude, $\mu_d$ is the
reduced mass of the decay particles and $p$ their relative momentum.
Therefore eq. (\ref{3part}) agrees with (\ref{invo})  when
we identify
\begin{equation}
\rho \left( m ,\Gamma \right) = \int d\Omega_p \mu_d |\vec p|
\mid D_\omega (m)
W(m,p) \mid ^2 \ = -\frac{1}{\pi}Im(D_\omega (m)),
\label{specdef}
\end{equation}where we used unitarity for the second identity.
 Eq. (\ref{specdef})
  agrees with the standard definition of a spectral function.

\section{Summary}

To summarize, we demonstrated that the interpretation of
the experimental results for the reaction $\pi^-p \to \omega n$
given in refs. \cite{Bin,Key,Kar} is incorrect. A
proper treatment of the independent variables leads to an
expression for the momentum dependence of the
integrated cross section that is consistent with
a constant matrix element near the production threshold.

The procedure of refs.  \cite{Bin,Key,Kar} was also used in ref. \cite{Wurz}
and thus our criticism applies to the conclusion
 of this paper as well. However, we want to emphasize that we regard the
 method of integrating
over the beam energy while keeping the final momentum fixed
 as useful way to examine the production of
narrow resonances close to their production threshold.
This technique
allows for a more direct access to the production amplitude and,
simultaneously, to an increase in the counting rate.

We demonstrated that  as the momentum $P^*$ grows, the formula for the
cross section reduces to the standard one
for the production of a stable particle,
as expected. The relevant parameter
is $\chi = \frac{2P^*\Delta P}{\mu\Gamma}$.

The knowledge of the transition amplitude $\pi N \to \omega N$
is an important input for several approaches investigating
$\omega$ production in hadron--hadron collisions
\cite{wil,kon,tobechecked,sib}.
A better understanding of its energy dependence therefore will
help us to get deeper insight in the strong interaction of vector
mesons and nucleons and nuclei in the intermediate energy regime.

{\bf Acknowledgments}

The authors are grateful to C.Wilkin for
useful discussion and information about the papers
\cite{Bin,Key,Kar}, and to K. Brinkmann, O. Krehl, N.N.Nikolaev, F.Plouin and
A. Sibirtsev for useful discussions. A.K. thanks V.Baru for
help in numerical estimations.
We also wish to thank J. Durso for editorial assistance.
A.K. acknowledges the hospitality of the Institute
f\"ur Theoretische  Kernphysik IKP, Forschungzentrum   J\"ulich.
A.K is also thankful to the Russian Foundation
for Basic Research (grant 98-02-17618) for partial support.
 C.H. is grateful for financial support by the COSY FFE--Project
No. 41324880, Department of Energy grant DE-FG03-97ER41014
and by the Humboldt Foundation.



%
%
%
\phantom{=}

%
%
%

\end{document}